\documentclass[aps,prd,twocolumn,nofootinbib,superscriptaddress]{revtex4-1}
\usepackage{graphicx}

\begin{document}

\pagestyle{plain}

\title{The Twin Higgs: Natural Electroweak Breaking from Mirror Symmetry}

\author{Z. Chacko}
\author{Hock-Seng Goh}
\affiliation{Department of Physics, University of Arizona, Tucson, AZ 85721}
\author{Roni Harnik}
\affiliation{
Department of Physics, University of California,
Berkeley, CA 94720 \\
Theoretical Physics Group,
Lawrence Berkeley National Laboratory, Berkeley, CA 94720}


\begin{abstract}

We present `twin Higgs models', simple realizations of the Higgs as a
pseudo-Goldstone boson that protect the weak scale from radiative corrections
up to scales of order 5 - 10 TeV. In the ultra-violet these theories have a
discrete symmetry which interchanges each Standard Model particle with a
corresponding particle which transforms under a twin or mirror Standard Model
gauge group.  In addition, the Higgs sector respects an approximate global
symmetry.  When this global symmetry is broken, the discrete symmetry tightly
constrains the form of corrections to the pseudo-Goldstone Higgs potential,
allowing natural electroweak symmetry breaking.  Precision electroweak
constraints are satisfied by construction.  These models demonstrate that,
contrary to the conventional wisdom, stabilizing the weak scale does not
require new light particles charged under the Standard Model gauge groups.

\end{abstract}

\pacs{} \maketitle



In the Standard Model (SM) the weak scale is unstable under quantum corrections. This
suggests the existence of new physics at or close to a TeV that protects the Higgs
mass parameter of the SM against radiative corrections. While the exact form that
such new physics takes is unknown there are several interesting alternatives. One
possibility, first proposed in~\cite{GP, KG} is that the Higgs is naturally light
because it is the pseudo-Goldstone boson of an approximate global symmetry. This idea
has recently experienced a revival in the form of little Higgs
theories~{\cite{Little1, Little2}} (for a clear review and more references see
{\cite{Review}})  that protect the Higgs mass from radiative corrections up to scales
of order 5 - 10 TeV.


In this paper we propose a class of simple alternative realizations of the Higgs as a
pseudo-Goldstone boson that also protect the weak scale from radiative corrections up
to scales of order 5 - 10 TeV.
In the ultra-violet these theories have a discrete
Z$_2$ symmetry which interchanges each Standard Model particle with a corresponding
particle which transforms under a twin or mirror Standard Model gauge group. In
addition, the
Higgs sector of the theory respects an approximate global $SU(4)$ symmetry.
Although the weak and electromagnetic interactions, as well as the top Yukawa
coupling, violate the global symmetry they all respect the discrete interchange
symmetry.  When $SU(4)$ is broken to $SU(3)$, the discrete symmetry tightly
constrains the form of corrections to the pseudo-Goldstone Higgs potential, allowing
natural electroweak symmetry breaking.

Although the smaller Yukawa couplings need not respect the discrete symmetry,
naturalness constrains the masses of most of the twin/mirror partners not to exceed a
few hundred GeV.  Precision electroweak constraints are satisfied by construction,
since although these new particles may be very light, they do not transform under the
SM gauge groups. This is in contrast to little Higgs theories where these constraints
are often a severe problem {\cite{Csaba}}.


We illustrate the basic idea by way of a simple example where the global
symmetry is realized linearly.  Consider a complex scalar field, $H$, that
transforms as a fundamental under a global $SU(4)$ symmetry. The potential
for this field is given by
\begin{equation}
\label{treepotential}
    V(H)= -m^2 H^\dagger H + {\lambda}{}(H^\dagger H)^2\,.
\end{equation}
Since the mass squared of $H$ is negative it will develop a VEV, $\langle |H|
\rangle = m/\sqrt{2\lambda}\equiv f$, that breaks $SU(4)\to SU(3)$ yielding 7
massless Nambu-Goldstone bosons.  We now break the $SU(4)$ explicitly by
gauging an $SU(2)_A\times SU(2)_B$ subgroup. The field $H$ transfoms
as~$\left({ H_A , H_B} \right)$ where $H_A$ is a doublet under $SU(2)_A$ and
$H_B$ is a doublet under $SU(2)_B$. At the end of the day we will identify
$SU(2)_A$ with $SU(2)_L$ of the SM.
Since $SU(4)$ is now broken explicitly, the would-be Goldstones
pick up a mass that is proportional to the explicit breaking.
Specifically, gauge loops contribute a quadratically divergent mass to the
components of $H$ as
\begin{equation}
    \Delta V= \frac{9 g_A^2 \Lambda^2}{64\pi^2}H_A^\dagger H_A
        + \frac{9 g_B^2 \Lambda^2}{64\pi^2}H_B^\dagger H_B +\ldots\,,
\end{equation}
a loop factor below the cutoff $\Lambda$ of the theory.  The
mechanism in our model hinges on the following simple observation. Suppose we
now impose an additional $Z_2$ symmetry, which we label `twin parity', which
interchanges $H_A$ and $H_B$ and also interchanges the gauge bosons of
$SU(2)_A$ with those of $SU(2)_B$.
This symmetry forces the two gauge couplings to be equal, $g_A=g_B\equiv g$.
The gauge contribution to the mass of $H$ is now
\begin{equation}
    \Delta V= \frac{9 g^2 \Lambda^2}{64\pi^2}
    (H_A^\dagger H_A+H_B^\dagger H_B)=
    \frac{9 g^2 \Lambda^2}{64\pi^2} H^\dagger H
\end{equation}
which is invariant under $SU(4)$ and therefore does not contribute a mass to the
Goldstones.  In other words, imposing twin parity constrains the
quadratically divergent mass terms to have an $SU(4)$ invariant form.
The Goldstones are therefore completely insensitive to quadratic divergences
from gauge loops.

Gauge loops will however contribute a logarithmically divergent term to the potential
that is not $SU(4)$ symmetric and has the general form $ \kappa \left(|H_A|^4 +
|H_B|^4 \right)$ where $\kappa$ is of order $g^4/16 \pi^2 {\rm log} \left(\Lambda/ gf
\right)$. Provided $\Lambda$ is not very much larger than $f$ this leads to the
would-be Goldstones acquiring a mass of order $g^2 f/4 \pi$ which is of order the
weak scale for $f$ of order a TeV.  Notice that we could have obtained exactly the
same result by imposing `mirror parity' - invariance under $t \rightarrow t$,
$\vec{x} \rightarrow - \vec{x}$ along with the interchange of every particle in
sector A with its CP conjugate in B.

At this point we note that the Higgs potential of Eq.~({\ref{treepotential}})
actually possesses a larger global $O(8)$ symmetry of which $U(4)$ is
merely a sub-group, and the 7 Goldstone bosons we have identified can also be
thought of as emerging from the breaking of $O(8)$ to $O(7)$. In
particular, this $O(8)$ symmetry includes the custodial $SU(2)$ of the Higgs
potential in the Standard Model.


This approach to stabilizing the weak scale against quantum corrections from gauge
loops can be generalized to include all the other interactions in the SM. To do this,
we gauge two copies of the SM, A and B, with our SM being SM$_{\rm A}$.
We can then extend the discrete symmetry in either of the following two ways:
1) Interchange every SM$_{\rm A}$ particle with the corresponding particle in 
SM$_{\rm B}$, or
2) Impose $t \rightarrow t$, $\vec{x} \rightarrow - \vec{x}$ along with the interchange 
of every SM$_{\rm A}$ particle with its CP conjugate in SM$_{\rm B}$.
These symmetries, while similar, are distinct. Each one relates the gauge and Yukawa
interactions in the A sector to those in the B sector. While the former is a simple
generalization of twin parity which we label `twin symmetry', the latter extends
mirror parity to the familiar mirror symmetry {\cite{mirror}}.
Either choice of the discrete symmetry ensures that any quadratically divergent
contribution to the Higgs mass has a form $\propto \Lambda^2 (|H_A|^2 + |H_B|^2)$
which is harmless due to its accidental $SU(4)$ symmetry. Although quantum
corrections to the quartic are in general not $SU(4)$ invariant, once again these
only lead to logarithmically divergent contributions to the mass of the
pseudo-Goldstone Higgs field, allowing for a natural hierarchy between $f$ and the
weak scale.

In both the twin and mirror symmetric cases the only renormalizable interactions
between the SM sector and the hidden sector allowed by gauge invariance are the Higgs
quartic, which is assumed to have an $SU(4)$ invariant form at the cut-off $\Lambda$,
and a mixing term between the hypercharge gauge boson and its partner, which we
neglect for the present discussion and will return to later.


At one loop the largest contribution to the pseudo-Goldstone Higgs potential arises
from the top Yukawa coupling, and is logarithmically sensitive to the cutoff.
However, in the twin symmetric case it is straightforward to make this
contribution finite.
One possible approach is to enlarge the approximate global symmetry of the
top Yukawa coupling to $SU(6)\times SU(4) \times U(1)$ with the $(SU(3)_c\times
SU(2)\times U(1))_{A,B}$ subgroups being gauged.
We do this by introducing the following chiral fermions
\begin{eqnarray}
    Q_L&=&(\mathbf{6}, \mathbf{\bar 4}) \nonumber\\
    &=& (\mathbf{3}, \mathbf{2} ;\mathbf{1}, \mathbf{1} ) +
    (\mathbf{1}, \mathbf{1} ;\mathbf{3}, \mathbf{2} ) +
    (\mathbf{3}, \mathbf{1} ;\mathbf{1}, \mathbf{2} ) +
    (\mathbf{1}, \mathbf{2} ;\mathbf{3}, \mathbf{1} ) \nonumber\\
    &\equiv& q_A + q_B + \tilde q_A + \tilde q_B\nonumber\\
    T_R&=&(\mathbf{\bar 6}, \mathbf{1})\nonumber\\
    &=&(\mathbf{\bar 3}, \mathbf{1} ;\mathbf{1}, \mathbf{1} ) +
    (\mathbf{1}, \mathbf{1} ;\mathbf{\bar 3}, \mathbf{1} )
    \equiv t_A + t_B
\end{eqnarray}
which transform as shown
under $SU(6)\times SU(4)$ and under $\left[SU(3)\times SU(2)\right]^2$,
where we have suppressed the hypercharge quantum numbers.  One can then write
an $SU(4)$ invariant Yukawa coupling
\begin{equation}
\label{topmodule}
    y H Q_L T_R + \mbox{h.c.}
\end{equation}
The $SU(4)$ symmetric matter content contains exotic left handed quarks,
$\tilde q_{A,B}$ that are charged under color of one sector and the weak
group of the twin, and vice versa. We introduce additional fermions with
opposite charge assignment, $\tilde q^c_{A,B}$ with which the exotic quarks
can get a~$Z_2$ symmetric mass
$M\left( \tilde q_A^c \tilde q_A +\tilde q_B^c \tilde q_B \right)\,. $
The mass parameter $M$ is the only source of $SU(4)$ breaking in the top
sector, and it only breaks this symmetry softly.
The top contribution to the Higgs potential in this model will then be
finite at one loop.

We now construct a realistic twin symmetric model that implements these symmetries
non-linearly.
The linear model we have been working with should be considered merely one
possibility for a UV-completion of the non-linear one, and others may well
exist.
The pseudo-Goldstone fields of the non-linear model are those which survive
after integrating out the radial mode of the field $H$ in the linear model.
We parametrize these degrees of freedom as
\begin{eqnarray}
\label{pseudo}
H = \exp(\frac{i}{f} {h}^a t^a) \pmatrix{ 0 \cr 0 \cr 0 \cr f }
\equiv \pmatrix{ 0 \cr 0 \cr 0 \cr f }
+ i \pmatrix{ h^1 \cr h^2 \cr h^3 \cr h^0 } + \ldots
\end{eqnarray}
where $h^{1,\ldots,3}$ are complex and $h^0$ is real.  In general the
effective theory for these fields will contain all of the operators allowed
by the non-linearly realized $SU(4)$ symmetry, suppressed by the cutoff scale
$\Lambda$.  However, in order to suppress custodial
SU(2) violation we assume that the symmetry which is non-linearly realized is in fact
$O(8)$. This provides additional restrictions on the form of the
interactions in the effective theory below $\Lambda$, allowing precision electroweak
constraints from higher dimensional operators to be naturally satisfied.
Assuming the theory is strongly coupled at the cutoff we can
estimate $\Lambda\sim 4 \pi f$.  
However, any potential for the
pseudo-Goldstone fields can only emerge from those interactions which violate
the global $O(8)$ symmetry, specifically their gauge and Yukawa couplings. In
particular the electroweak gauge interactions and the top Yukawa contribute
the most to the pseudo-Goldstone potential and must therefore be studied in
detail. We will thus calculate the contributions to the one loop
Coleman-Weinberg (CW) potential~\cite{CW} from these couplings. At one loop
the gauge and top sectors contribute separately, simplifying the
calculation.

As before, we gauge two copies of the SM, A and B.
The vev $f$
breaks $SU(2)_B\times U(1)_B$ down to a single $U(1)$, giving $W_B$ and $Z_B$
masses of order $gf$.
The $SU(2)_A$ doublet $h^T\equiv (h^1, h^2)$ is left uneaten and
is identified as the SM Higgs. The couplings of the pseudo-Goldstone
fields to the $SU(2) \times U(1)$ gauge fields and their mirror partners are
given by expanding out $H = (H_A, H_B)$ in terms of the pseudo-Goldstones as
given by eq. ({\ref{pseudo}}) in the interaction
\begin{equation}
\left|\left(\partial_\mu + ig W_{\mu, A}
+ \frac{i}{2} g' B_{\mu, A}\right)H_A \right|^2
+ \left( A \rightarrow B \right)\,.
\end{equation}
A simple way of calculating the effective potential is to calculate the
vacuum energy as a function of the field dependent masses of all of the
fields in the theory.
In the absence of
quadratic divergences this leads to the formula
\begin{equation}
    V_{CW} = \pm \frac{1}{64\pi^2} \sum_i {M_i^4}
    \left(\log\frac{\Lambda^2}{M_i^2} + \frac{3}{2}\right)
\end{equation}
where the sum is over all degrees of freedom, the sign being negative for
bosons and positive for fermions. In evaluating this sum the higher order
terms in the expansion eq. ({\ref{pseudo}}) are often useful, giving
\begin{eqnarray}
        \label{expand}
        H_A^\dagger H_A &=& h^\dagger h - \frac{(h^\dagger h)^2}{3f^2} +\ldots
        \nonumber\\
        H_B^\dagger H_B &=& f^2 - h^\dagger h + \frac{(h^\dagger h)^2}{3f^2}
        -\ldots
\end{eqnarray}
This expansion manifests the fact that the $Z_2$ invariant $(|H_A|^2 +
|H_B|^2)$ is independent of the Higgs.
Writing the Higgs potential in the form
\begin{equation}
\label{potential}
        V(h)= m_h^2 h^\dagger h + \lambda_h (h^\dagger h)^2 + \ldots
\end{equation}
we find that the contribution to the Higgs mass term from the gauge sector is
\begin{eqnarray}
\label{gauge}
        m_h^2
    &=& \frac{6 g^2 M^2_{W_B}}{64\pi^2 }
    \left( \log\frac{ \Lambda^2} {M^2_{W_B}} + 1 \right) \\
    &+& \frac{3 (g^2+g'^2) M^2_{Z_B}}{64\pi^2 }
    \left( \log\frac{ \Lambda^2} {M^2_{Z_B}} + 1 \right) \,
    \nonumber
\end{eqnarray}
where
$M_{W_B}^2= g^2 f^2/2$ and $M_{Z_B}^2= (g^2+g'^2)f^2/2 \,$.
Eq.~(\ref{gauge}) holds if electromagnetism in the twin sector is an unbroken
gauge symmetry as in the SM. However it is also possible that QED
in the twin sector is a broken symmetry and that the twin photon has a
mass.  This could arise if, for example, the hypercharge gauge boson in the
twin sector has a mass $M_{B}$ which softly breaks the twin symmetry.
We do not specify a dynamical origin for this mass since it is technically
natural for the dynamics which generate it to lie at scales above the
cutoff $\Lambda$. In the limit that $M^2_B \gg g'^2 f^2$ the second term in
eq.~({\ref{gauge}}) becomes approximately
\begin{equation}
 \frac{3
  g^2 M^2_{W_B}}{64\pi^2 }
        \left( \log\frac{ \Lambda^2} {M^2_{W_B}} +1 \right) +
        \frac{3 g'^2 M^2_{B}}{64 \pi^2}
        \left( \log\frac{\Lambda^2}{M_{B}^2} +1 \right)
\end{equation}
The contribution to the Higgs quartic from this sector is small and can be neglected.

We now turn to the top sector. The couplings of the pseudo-Goldstone fields
to the top quark are obtained by expanding out $H$ as in eq. ({\ref{pseudo}})
in the $SU(4) \times SU(6)$ invariant interaction ($y H Q_L T_R +
\mbox{h.c.}$) of eq.  ({\ref{topmodule}}). The $h$ dependent masses of the
fields in the top sector are determined from this and from the $SU(4)$ breaking
mass term, and can be expressed as
\begin{eqnarray}
       m^2_{t_A}=\frac{y^2 M^2}{M^2 +y^2f^2} h^\dagger h &\qquad&
       m^2_{T_A}= M^2 +y^2f^2 \nonumber\\
       m^2_{t_B}= y^2 f^2 \qquad\qquad &\qquad&
       m^2_{T_B}= M^2
\end{eqnarray}
to leading order in $|h|^2$, where we have asumed for simplicity that $y$ is real.
This leads to the following contributions to the Higgs potential of
eq.~({\ref{potential}}).
\begin{eqnarray}
        m_h^2&=&
        \frac{3}{8\pi^2}\frac{y^2 M^2}{M^2-y^2f^2}
        \left( M^2\log\frac{m_{T_A}^2}{m^2_{T_B}}
        -y^2f^2 \log\frac{m^2_{T_A}}{m^2_{t_B}}\right)\,,
        \nonumber\\
        \lambda_h &=& -\frac{m_h^2}{3f^2} +
        \frac{3}{16 \pi^2}\frac{y^4 M^4}{(M^2+y^2f^2)^2}
        \log\frac{m^2_{T_A}}{m^2_{t_A}} \nonumber\\
        &+& \frac{3}{16\pi^2}
        \frac{y^4M^4 (M^2+y^2f^2)}{(M^2-y^2f^2)^3}
        \log\frac{m^2_{T_B}}{m^2_{t_B}} \nonumber\\
        &-& \frac{3}{32\pi^2}
        \left[\frac{4 y^4 M^4}{(M^2-y^2f^2)^2}
        +\frac{y^4 M^4}{(M^2+y^2f^2)^2}\right]
\end{eqnarray}

In order to generate a mild hierarchy $\langle h \rangle < f$ so that in the strong
coupling limit the cutoff $\Lambda ~ 4 \pi f$ is of order 5 TeV we add to the theory a
`$\mu$ term' that softly breaks the discrete $Z_2$ twin symmetry. This term takes the
form $\mu^2 H_A^\dagger H_A$ and contributes to $m_h^2$ and $\lambda_h$ as dictated by
eq. ({\ref{expand}}). In addition, since the smaller Yukawa couplings do not contribute
significantly to the Higgs potential, we do not require them to respect the discrete
symmetry. In this non-linear model, the absence of quadratically divergent
contributions to the Higgs mass can be understood as a consequence of cancellations
between the familiar SM loop corrections and new loop corrections that arise from the
(mostly non-renormalizable) couplings of the Higgs to the twin sector.

For phenomenological purposes we divide twin symmetric models into two classes -
those where the top sector is extended as in eq.~({\ref{topmodule}}), and those where
it is not.  As we now explain, the experimental constraints in these two cases are
different. In the first case the exotic quarks $\tilde q_{A,B}$ and $\tilde
q^c_{A,B}$, which are charged under both $U(1)_A$ and $U(1)_B$, lead to kinetic
mixing between the photon and its twin partner at one loop {\cite{Holdom}}. Since the
experimental constraints on such mixing are very severe the twin photon must be
heavy. In the second case, however, there are no particles charged under both sets of
gauge groups, and a preliminary analysis does not reveal any non-zero contribution to
the kinetic mixing term up to three loop order. In this scenario it may therefore be
phenomenologically allowed for the twin photon to be massless, provided a kinetic
mixing term is not present at the cutoff. The mirror symmetric model shares the same
phenomenology as the twin symmetric model without the extended top sector.

We now study each of these two scenarios in more detail, starting with theories with
the extended top sector. In this case the strongest bound arises from the requirement
that the twin neutrinos (and the twin photon itself) not contribute significantly to
the energy density of the universe at the time of Big Bang
Nucleosynthesis~(BBN)~{\cite{CG,Davidson}}. This constraint can be satisfied if the
following two conditions are met,
there is large
entropy production during the QCD phase transition, significantly more 
than during the corresponding
transition in the twin sector, and  
The two sectors are not in thermal equilibrium
at any time after the QCD phase transition.  
Since the dynamics of the QCD phase transition is expected to be sensitive to the
number of light quarks and their masses, which are not constrained to be the same in
the two sectors, it is certainly plausible that the first condition is satisfied.
What about the second? If the mixing term is zero at the cutoff and is only generated
at the one-loop level through the exchange of the exotic quarks, the two sectors will
not be in equilibrium below a few hundred MeV provided the twin photon mass $M_{B}$
is larger than a few hundred GeV. In such a scenario the twin electron cannot go out 
of
the bath by annihillating into photons once the temperature falls below it's mass,
as in the SM. Instead the twin electron must be extremely light so as not to
contribute too much to the energy density of the universe at late times. We expect
that twin baryons will constitute some or all of the dark matter in the universe,
depending on the baryon asymmetry in the mirror sector.

Although this model predicts the existence of new light twin states, the fact that
these particles are not charged under the SM gauge group implies that it may not be
easy to test. In particular, precision electroweak constraints are easily satisfied.
However, one possibility is to look for invisible decays of the SM Higgs into twin
fermions {\cite{Foot}}. The relevant vertex arises from substituting the expansion
eq. ({\ref{pseudo}}) into the Yukawa coupling of $H_B$ to twin fermions.  The
branching ratio for invisible Higgs decays is of order $|\langle{h}\rangle/f|^2$.


We now estimate the fine-tuning in this class of models for two sets of
parameters.  For $f = 800$ GeV, $\Lambda \sim 4 \pi f \approx 10$ TeV, $M =
6.0$ TeV, $M_B = 1$ TeV we find that in order to obtain the SM
values of $M_W$ and $M_Z$ we need the soft $Z_2$ breaking parameter $\mu
\approx$ 240 GeV. The Higgs mass is then about 120 GeV. Estimating the fine-tuning
as $\partial \; {\rm log} M_Z^2 / \partial \; {\rm log} \mu^2 $ we find that
it is of order 13 \% (1 in 8). For $f = 500$ GeV, $\Lambda \sim 4 \pi f
\approx 6$ TeV, $M = 5.5$ TeV, $M_B = 1$ TeV we find the soft $Z_2$ breaking
parameter $\mu$ needs to be around 150 GeV.  The Higgs mass is again around
120 GeV and the fine-tuning 38\% (1 in 3).
This shows that these models stabilize the weak scale up to 5-10 TeV.

Let us now turn to mirror symmetric models (and twin symmetric models without the
extended top sector). We are specifically interested in the scenario where the mirror
photon is massless, since it appears current experimental bounds cannot exclude this
possibility. This class of models also predict new light mirror fermions. These may
now have tiny fractional electric charges if the kinetic mixing term between the
photon and its mirror partner is very small but non-zero.  Apart from this, the
phenomenological implications are expected to be similar to the case with the
extended top sector.  We now estimate the fine-tuning for one specific parameter
choice. Note that the formulas of the previous section generalize to the case without
the extended top sector when the limit $M \rightarrow \Lambda$ is taken, up to finite
terms.  For $f = 800$ GeV, $\Lambda \sim 4 \pi f \approx 10$ TeV we find that the
Higgs mass is 166 GeV and the fine-tuning is $\sim$11\% (1 in 9). For $f= 500$ GeV,
$\Lambda\sim 4\pi f \approx 6$ TeV we get a Higgs mass of 153 GeV with a fine tuning
of 31\% (1 in 3). This shows that this class of models also stabilizes the weak scale
up to 5-10 TeV. 


In summary we have constructed a new class of models where the Higgs emerges
as a pseudo-Goldstone whose mass is protected against radiative corrections
up to scales of order 5-10 TeV. These theories demonstrate that, contrary to
the conventional wisdom, stabilizing the weak scale does not require new light
particles transforming under the SM gauge groups.
\\

\noindent
{\bf Acknowledgments --} 
Z.C thanks M. Luty, N. Okada and W.D. Toussaint for discussions. Z.C and H.S.G are
supported by the NSF under grant PHY-0408954.  R.H is supported in part by the DOE
under contract DE-AC03-76SF00098 and in part by NSF grant PHY-0098840.

\end{document}